\documentclass[aps,prd,amsmath,amssymb,twocolumn,preprintnumbers,nofootinbib]{revtex4}
\pdfoutput=1
\usepackage{graphicx}
\usepackage{multirow}
\usepackage{color}
\newcommand{\beq}{\begin{eqnarray}}
\newcommand{\eeq}{\end{eqnarray}}

\newcommand{\bmp}{\noindent\begin{minipage}{16cm}}
\newcommand{\emp}{\end{minipage}\vskip 7mm} 

\usepackage{dcolumn}
\usepackage{bm}
\usepackage{bbm}
\usepackage{pxfonts}
\usepackage[margin=5pt, font=normalsize,labelfont=bf,format=plain,justification=centerlast]{caption}
\usepackage{subfig}

\definecolor{rossoCP3}{cmyk}{0,.88,.77,.40}

\def\lsim{\mathrel{\rlap{\lower4pt\hbox{\hskip1pt$\sim$}}
    \raise1pt\hbox{$<$}}}                
\def\gsim{\mathrel{\rlap{\lower4pt\hbox{\hskip1pt$\sim$}}
    \raise1pt\hbox{$>$}}}                

\newcommand{\be}{\begin{eqnarray}}
\newcommand{\ee}{\end{eqnarray}}

\usepackage{fancyhdr}
\begin{document}
\title{\Large  \color{rossoCP3} UV and IR Zeros of Gauge Theories \\ at \\The Four Loop Order and Beyond}
\author{Claudio {\sc Pica}$^{\color{rossoCP3}{\varheartsuit}}$}
\email{pica@cp3.sdu.dk} 
\author{Francesco {\sc Sannino}$^{\color{rossoCP3}{\varheartsuit}}$}
\email{sannino@cp3.sdu.dk} 
\affiliation{
$^{\color{rossoCP3}{\varheartsuit}}${ CP}$^{ \bf 3}${-Origins}, 
Campusvej 55, DK-5230 Odense M, Denmark.}
\begin{abstract}
We unveil the general features of the phase diagram for any gauge theory with  fermions transforming according to distinct representations of the underlying gauge group, at the four-loop order. We classify and analyze the zeros of the perturbative beta function and discover the existence of a rich phase diagram. The anomalous dimension of the fermion masses, at the infrared stable fixed point, are presented. We show that the infrared fixed point, and associated anomalous dimension, are well described by the all-orders beta function for any theory. We also argue the possible existence, to all orders, of a nontrivial ultraviolet fixed point for any non-Abelian gauge theory at large number of flavors.  \\[.1cm]
{\footnotesize  \it Preprint: CP$^3$-Origins-2010-51}
\end{abstract}

\maketitle
\thispagestyle{fancy}

Determining the phase structure of generic gauge theories of fundamental interactions is crucial in order to be able to select relevant extensions of the standard model of particle interactions \cite{Sannino:2009za}. In particular we are interested in elucidating the physics of non-Abelian gauge theories as function of the number of flavors, colors and matter representation.  

To gain a quantitative analytic understanding of the phase structure of different gauge theories we investigate the zeros of the  perturbative beta function to the maximum known  order   and for one of the zeros also the limit of large number of flavors   to all-orders.

\section{Zerology}

We consider the perturbative expression of the beta function and the fermion mass anomalous dimension for a generic gauge theory with only fermionic matter in the $\overline{\rm MS}$ scheme to four loops which was derived in \cite{vanRitbergen:1997va,Vermaseren:1997fq}: 
\begin{eqnarray}
\label{beta}
\frac{d a }{d \ln \mu^2}  & = &
\beta(a) = -\beta_0 a^2 - \beta_1 a^3
-\beta_2 a^4
-\beta_3 a^5 + O(a^6)\ , \hspace{.5cm}
\end{eqnarray}
\begin{eqnarray}
 \label{gamma}
-\frac{d\ln m }{ d \ln \mu^2}  & = & 
\frac{\gamma (a)}{2}  =  \gamma_0 a +  \gamma_1 a^2
+ \gamma_2 a^3
+ \gamma_3 a^4 + O(a^5) \ ,\hspace{.5cm}
\end{eqnarray}
where $m=m(\mu^2)$ is the renormalized (running)  fermion mass and
$\mu$ is  the renormalization point in the $\overline{\rm MS}$ scheme and
$a=\alpha/4\pi=g^2/16\pi^2$ where $g=g(\mu^2)$ is the renormalized 
coupling constant  of the theory.

The explicit expression of the coefficients above are reported in the appendix \ref{beta-a} for completeness. 
 Note also that the beta function is gauge independent, order by order in perturbation theory \cite{vanRitbergen:1997va}. 
 The same also holds for the anomalous dimension of the fermion mass $\gamma$. 

\begin{figure*}[ht!]
	\subfloat[First kind of topology.]{\label{1a} \includegraphics[width=0.45\textwidth]{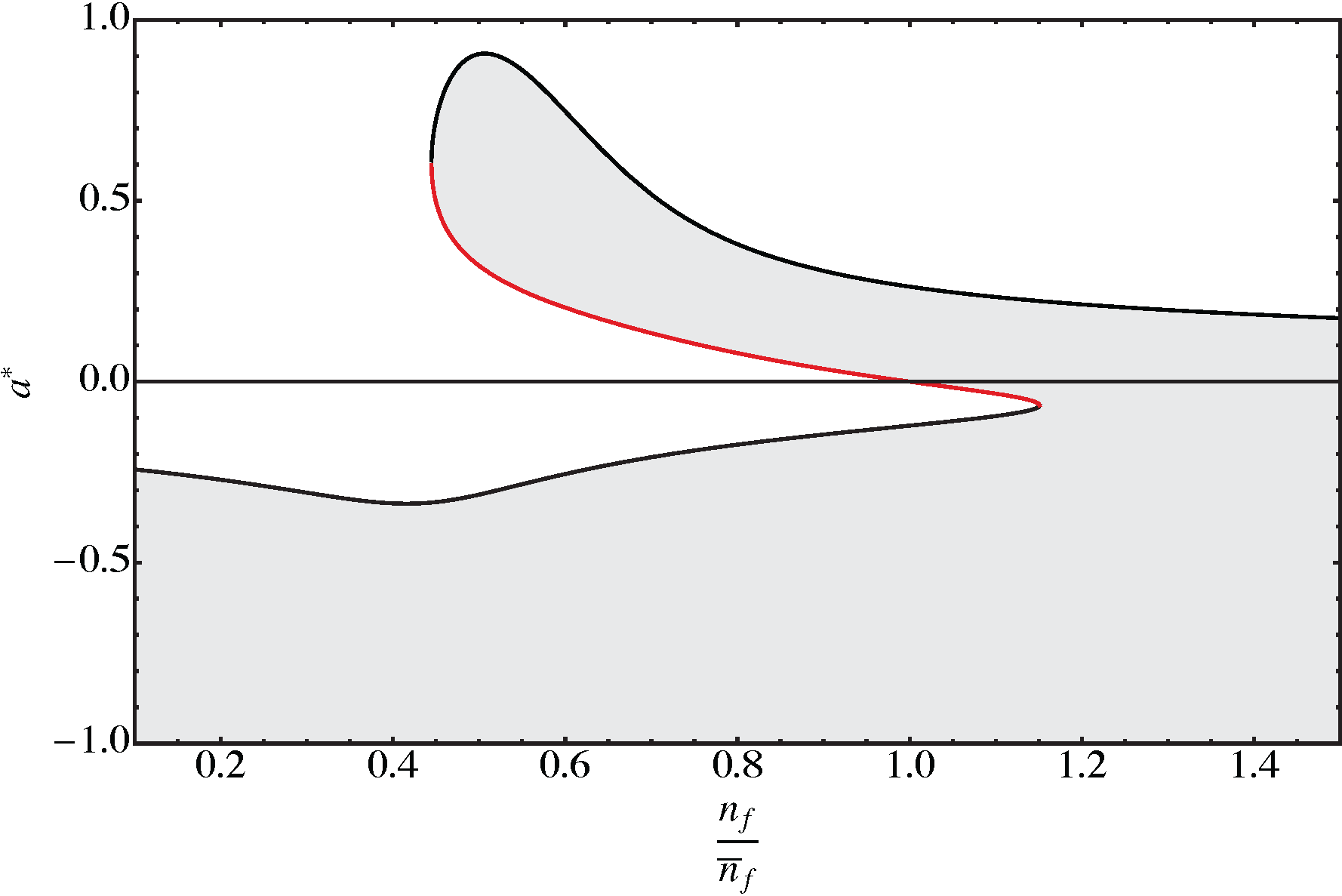}}
	\hfill
	\subfloat[Second kind of topology.]{\label{1b} \includegraphics[width=0.45\textwidth]{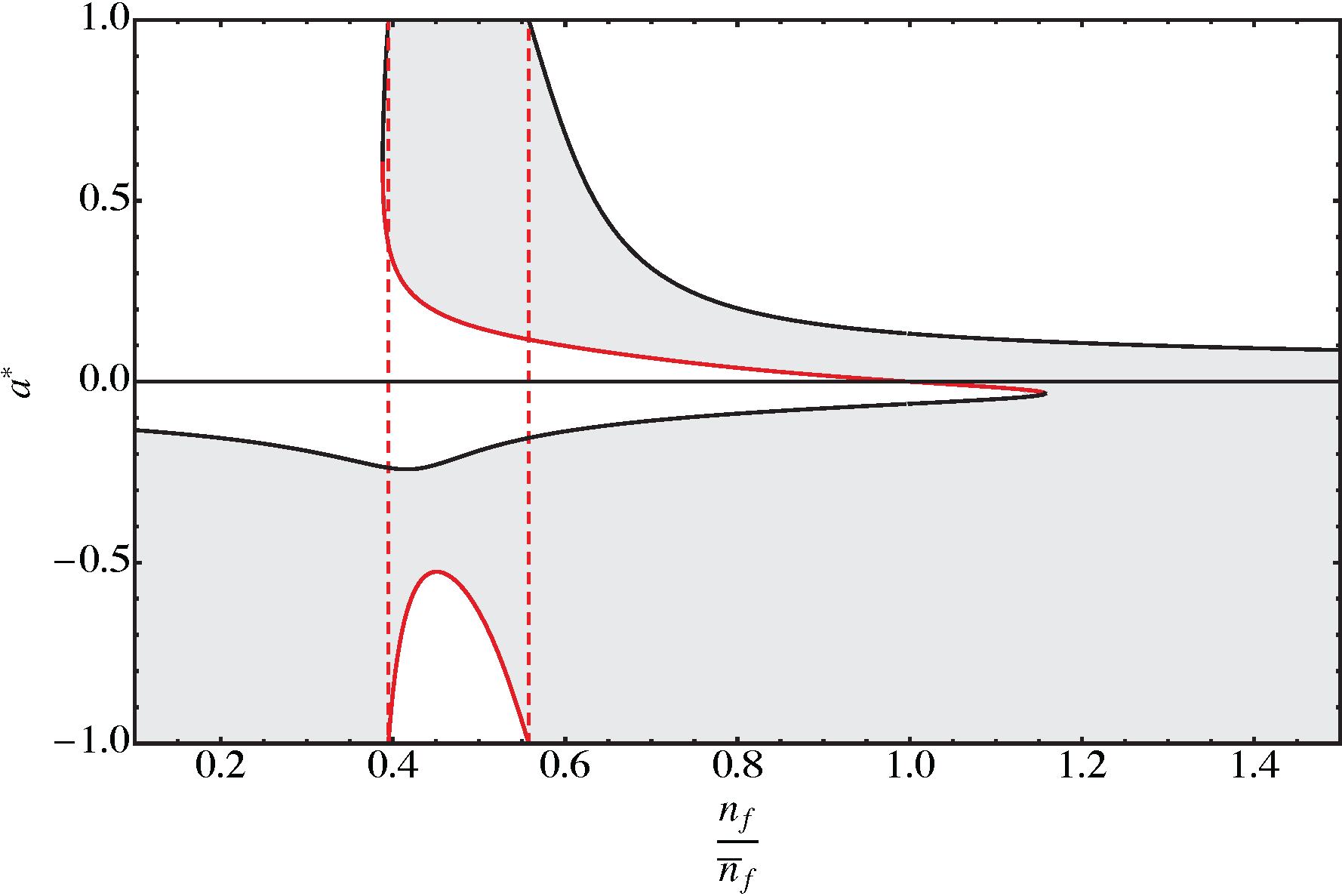}}\\
	\subfloat[Third kind of topology.]{\label{1c} \includegraphics[width=0.45\textwidth]{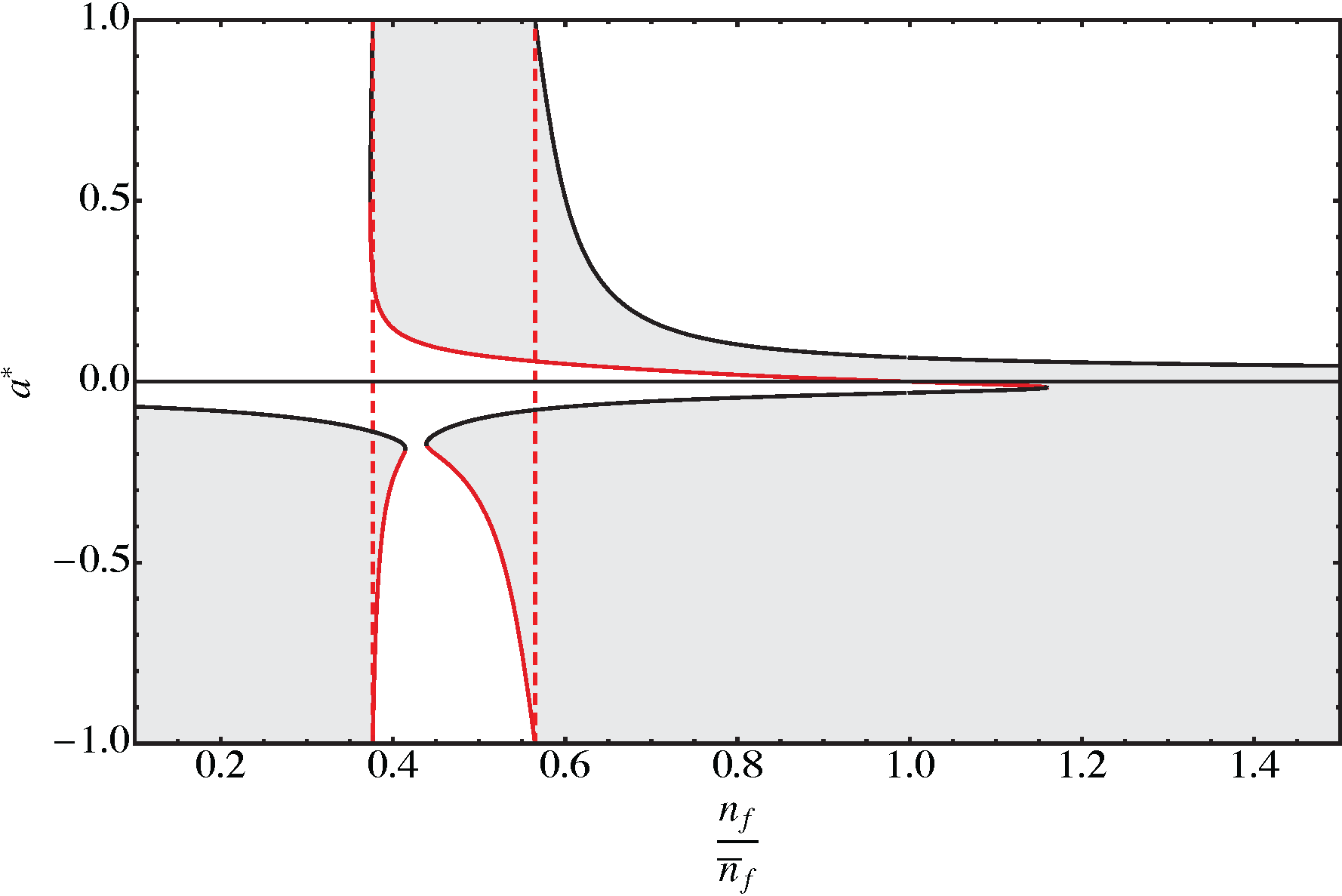}}
	\hfill
	\subfloat[Fourth kind of topology.]{\label{1d} \includegraphics[width=0.45\textwidth]{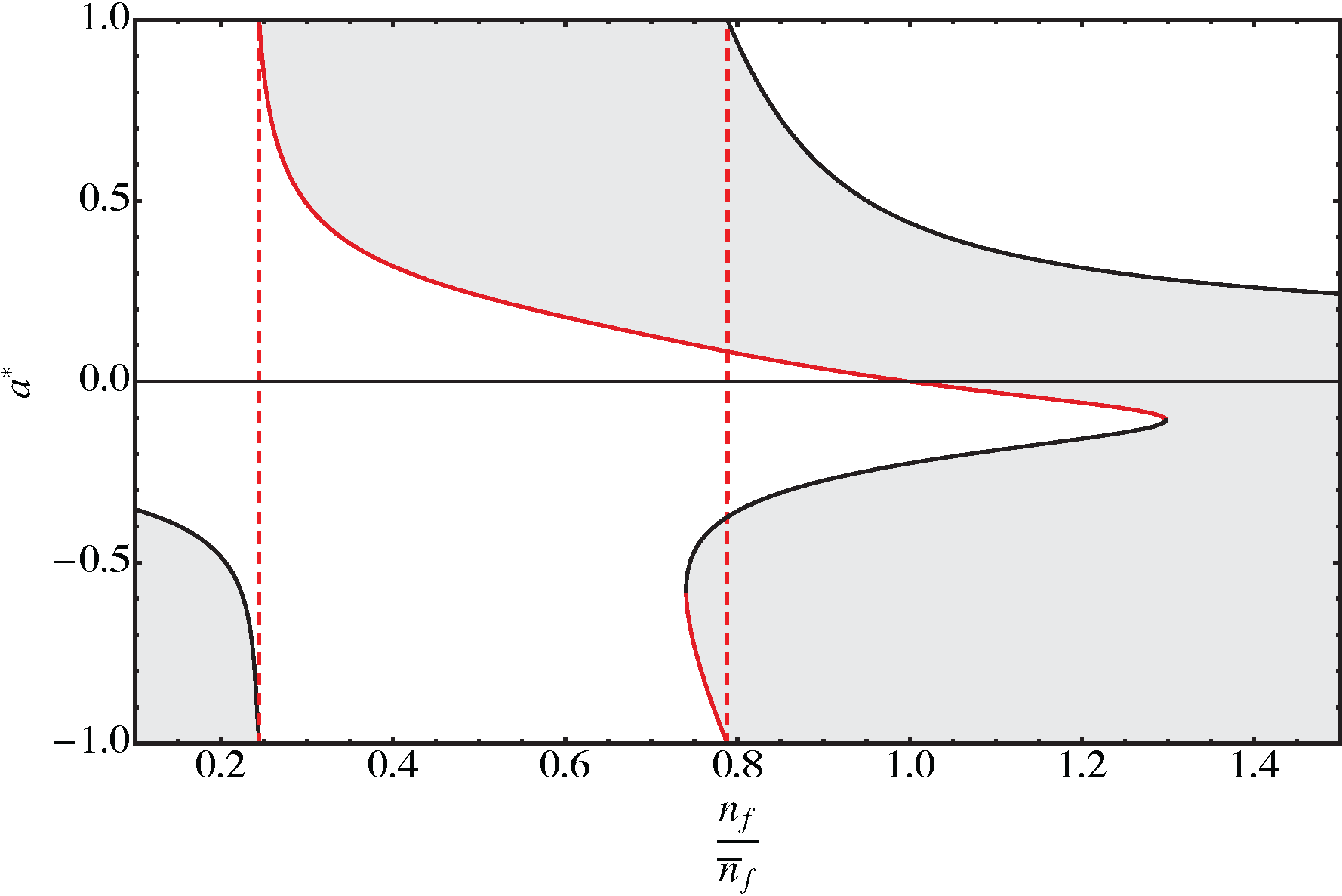}}
		\caption{The four different topologies displayed above classify the entire {\it zerology} landscape. We show, in each plot, the regions of positive (gray) and negative (white) values of the beta function for different gauge theories. The solid lines, per each figure, are the locations of the zeros of the beta functions. The lines of UV fixed points are in black while the IR ones in red. We have defined $\displaystyle{a^{\ast} = \frac{2}{\pi}\arctan \left( 5 a\right)}$. The vertical dashed red-lines correspond to the location where one zero approaches infinity. }
	\label{zerology}
\end{figure*}

Here we investigate the structure of the zeros of the four-loops beta function for any matter representation and gauge group. Interestingly we find a {\it universal} classification of the behavior of the  zeros as function of the number of flavors $n_f$.

The first observation is that due to the fact that the beta function is a polynomial of degree five in $\alpha$ there are five complex zeros. Since one can factor out $\alpha^2$,  the beta function will always have a double zero at the origin. The other three zeros determine the interesting properties of the theory, to this loop order. In the following we will focus on these three zeros which can be either all real or one real and two complex.

To elucidate the landscape of possible topologies we plot the real nontrivial zeros as function of the number of flavors normalized to the one above which asymptotic freedom is lost ($\overline{n}_f$) in Fig.~\ref{zerology}. Solid black lines correspond to the location of the ultraviolet (UV) zeros while red-ones to the infrared (IR) stable fixed points.  The shaded areas denote the regions where the $\beta$ function is positive. We have rescaled the vertical axis using the function $\displaystyle{a^{\ast} = 2\arctan \left( 5 a\right)}/\pi$,  mapping $[-\infty, +\infty]$  into the interval $[-1,1]$.

The best way to read these figures is to imagine a straight vertical line corresponding to a fixed value of $n_f$. The intersection of this line with the solid curve determines the number of the zeros, the color of the curves the type of zeros (if red is IR and if black is UV), and finally the corresponding horizontal value is the coupling location. We term the landscape of the zeros the {\it zerology}.   

We   investigate also the negative values of $\alpha$ since this is the most natural mathematical setting. In fact, the properties of the pure Yang-Mills theory at negative $\alpha$ have already been studied on the Lattice by Li and Meurice in \cite{Li:2004bw}.  There the authors have shown interesting relations between the positive and negative regions of  $\alpha$. 
 
It is possible to identify  four distinct topologies able to fully represent the entire zerology for any gauge group and matter representation which are reported in Fig.~\ref{zerology}. 

 We start by summarizing a number of general features that we have identified:
\begin{itemize}
\item At small number of flavors there is only a negative ultraviolet zero.  

\item At around and above $\overline{n}_f$ we observe the existence of three zeros, two ultraviolets and one infrared. The infrared one, near $\overline{n}_f$, is the Banks-Zaks \cite{Banks:1981nn} point.
Above $\overline{n}_f $, the IR fixed point is now at a negative values of $\alpha$ and at a new critical number of flavors collides with the UV fixed point zero at a negative value of the coupling, forming a double zero. At this point the beta function is positive for any negative alpha. 

\item At very large number of flavors the UV fixed point, for positive values of alpha, always exists and approaches zero asymptotically as $n_f^{-{2}/{3}}$.  The explicit derivation is provided in Sec.~\ref{salvo}.

\item By increasing $n_f$ from zero there is always a critical number of flavors above which an IR fixed point emerges for positive $\alpha$.

\end{itemize}
The distinguishing feature of different topologies is how the zeros merge or disappear as function of $n_f$. 

The topology A (Fig.~\ref{1a}) is characterized by the fact that the zeros always  remain at finite values of the coupling. This means that when a zero disappear it has to annihilate with another one. This happens at two distinct locations. One at a positive value of the coupling  and the other at a negative one occurring after asymptotic freedom is lost.

In the topology B, represented in Fig.~\ref{1b}, as for the previous case, we still observe the merging of the IR and UV zeros at two different number of flavors. In this case, however, there is a region in the number of flavors, where the UV fixed point located at positive couplings reaches infinity at finite $n_f$ and appears on the negative axis as an IR fixed point. The region where the new IR fixed point appears (on the negative coupling constant axis) ends before asymptotic freedom is lost.

The defining feature shown in Fig.~\ref{1c} for topology C is that the appearance of two more merging points on the negative axis.

In Fig.~\ref{1d}, topology D,  one observes that the IR zero at a positive value of the coupling reachers infinity at a finite value of the number of flavors, which is the distinctive feature of this topology.

A new feature at the four-loop order is that two positive nontrivial zeros, one IR and the other UV, can emerge simultaneously and can annihilate at a particular value of $n_f$.   At the two-loop level this feature does not exist and, in particular, no nontrivial ultraviolet fixed point is seen. 

As an example where these topologies arise we consider $SU(N)$ with fundamental fermions as function of $N$. {}For $N=2$ and $3$ the topology A occurs. Increasing $N$ the maximum value reached by the positive UV zero  increases and for $N=4$  it reaches infinity and therefore it enters topology B. Increasing $N$ further the local maximum of the IR negative zero-curve increases till it pinches the UV negative zero line for $N=11$ entering topology C. Topology $D$ is not realized in this case. On the other hand any $SU(N)$ gauge theory with $N \geq 2$ fermions and fermions in the adjoint representation lead to topology $D$. 

In table~\ref{topotable}  we catalogue the four-loop zerology for $SU(N)$, $SO(N)$ and $SP(2N)$ gauge theories with fermions transforming according to the fundamental and  the 2-index representations. 
\begin{table}
\begin{tabular}{c|c|c|c|c}
Rep. &Top. A&Top. B&Top. C&Top. D\\\hline \hline
\multicolumn{5}{c}{$SU(N)$}\\\hline \hline
FUND& $N=2,3$ & $4\leq N \leq 11$  & $N\geq 12$ & -- \\\hline
ADJ& -- & -- & -- & $N\geq 2$ \\\hline
2-SYM& -- & -- & -- & $N\geq2$ \\\hline
2-ASY& $N=3,4,5$ & $N=6,7$ & $8\leq N\leq 26$ & $N\geq 27$ \\\hline \hline
\multicolumn{5}{c}{$SO(N)$}\\\hline \hline
FUND&  $4\leq N\leq 8$ & $9\leq N \leq 22$ & $N\geq 23$ &$N=3$\\\hline
ADJ& --& -- & -- &$N\geq3$\\\hline
2-SYM& --& -- & -- &$N\geq3$\\\hline \hline
\multicolumn{5}{c}{$SP(2N)$}\\\hline \hline
FUND& $1\leq N \leq 2$ & $3\leq N\leq 9$ & $N\geq 10$ & -- \\\hline
ADJ& --& -- & -- &$N\geq1$ \\\hline
2-ASY& $2\leq N \leq 3$& $N=4$ & $5\leq N\leq 13$ & $N\geq 14$\\\hline
\end{tabular}
\caption{Catalogue of the four-loop  zerology for $SU(N)$, $SO(N)$ and $SP(2N)$ gauge theories with fermions transforming according to the fundamental and  the 2-index representations.\label{topotable}}
\end{table}


\section{Conformal Window}
The conformal window is defined as the region in theory space, as function of number of flavors and colors where the underlying gauge theory displays large distance conformality for a positive value of the coupling $\alpha$.  $\overline{n}_f$ constitutes the upper boundary of the conformal window and the lower boundary here is estimated by identifying  for which number of flavors the theory looses the infrared fixed point at a given number of colors. Due to the fact that we are using a truncated beta function the true window will be quantitatively different. 

\begin{figure*}[p]
\begin{center}
\includegraphics[width=.50\textwidth]{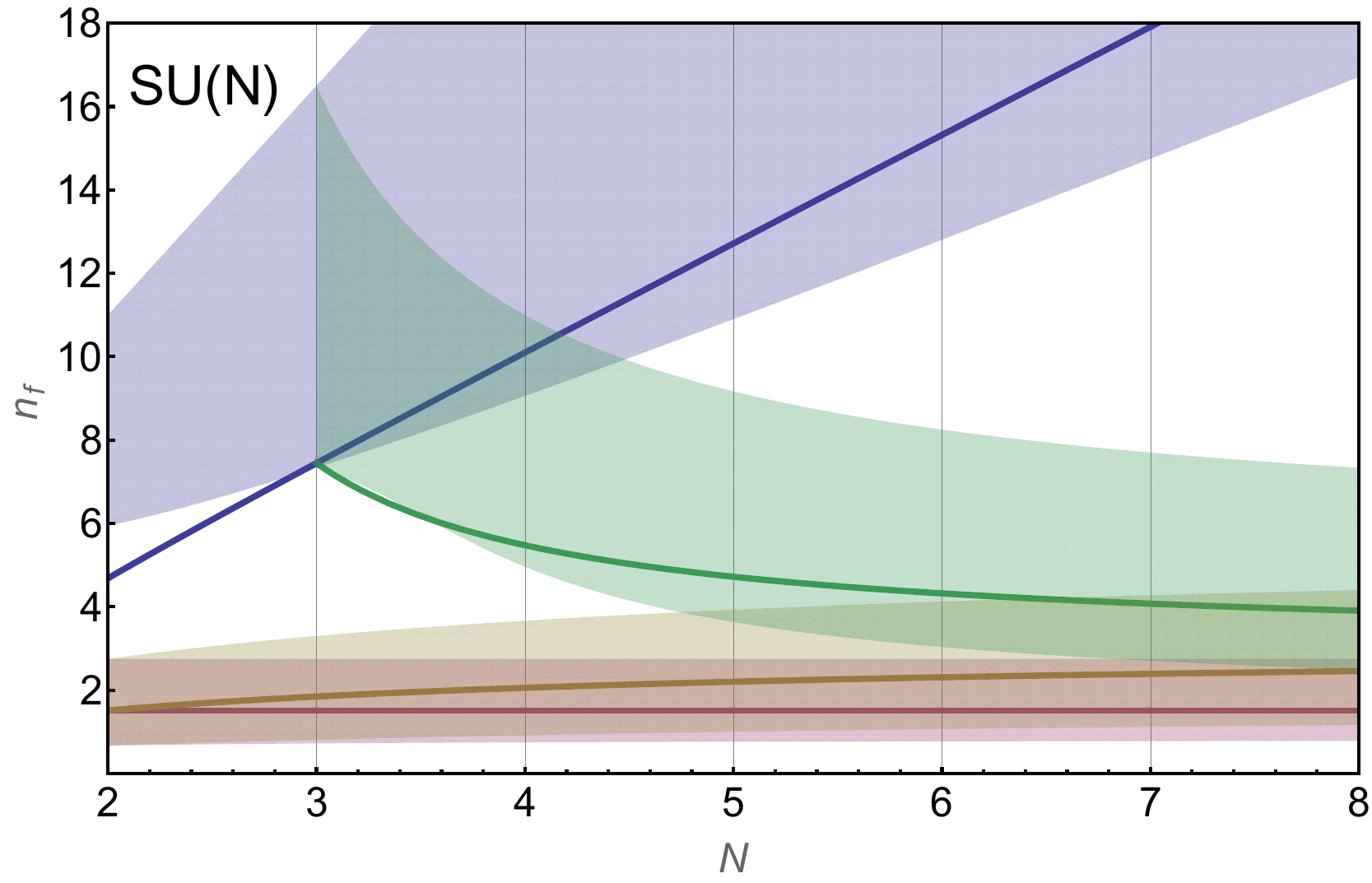}  
\caption{Conformal window for $SU(N)$ groups for the fundamental representation (upper light-blue), two-index antisymmetric (next to the highest light-green), two-index symmetric (third window from the top light-brown) and finally the adjoint representation (bottom light-pink). The lower boundary corresponds to the point where the infrared fixed point disappears at four loops. The solid thick lines correspond to the number of flavors for which the all-orders beta function predicts an anomalous dimension equal to unity.}    
\label{4loop-PD}
\includegraphics[width=.50\textwidth]{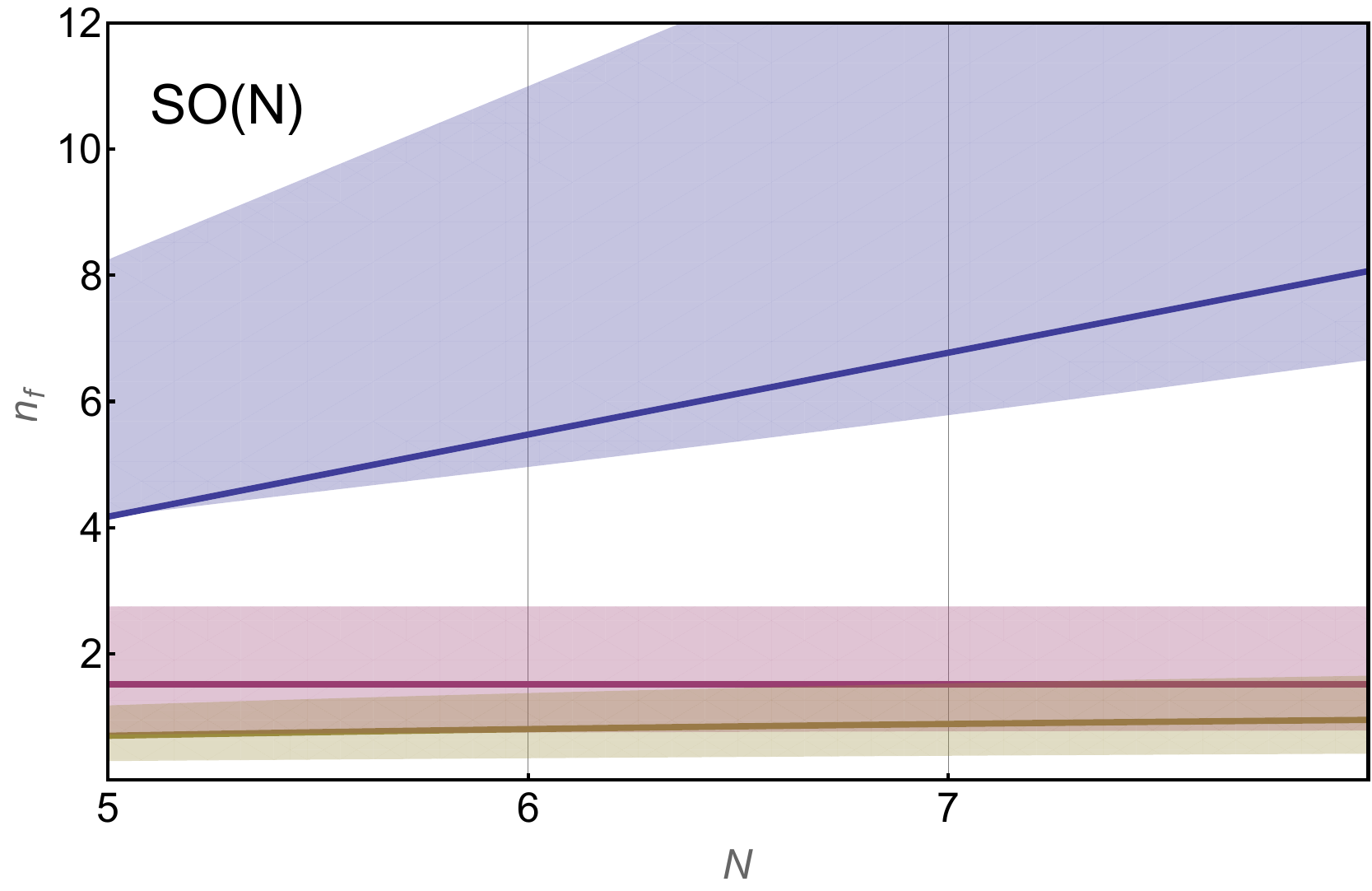}  
\caption{Conformal window for $SO(N)$ groups for the fundamental representation (upper light-blue), two-index antisymmetric (which is the adjoint and second from the top (pink-region)), two-index symmetric (bottom window in light-brown).}    
\label{SO-4loop-PD}
\includegraphics[width=.50\textwidth]{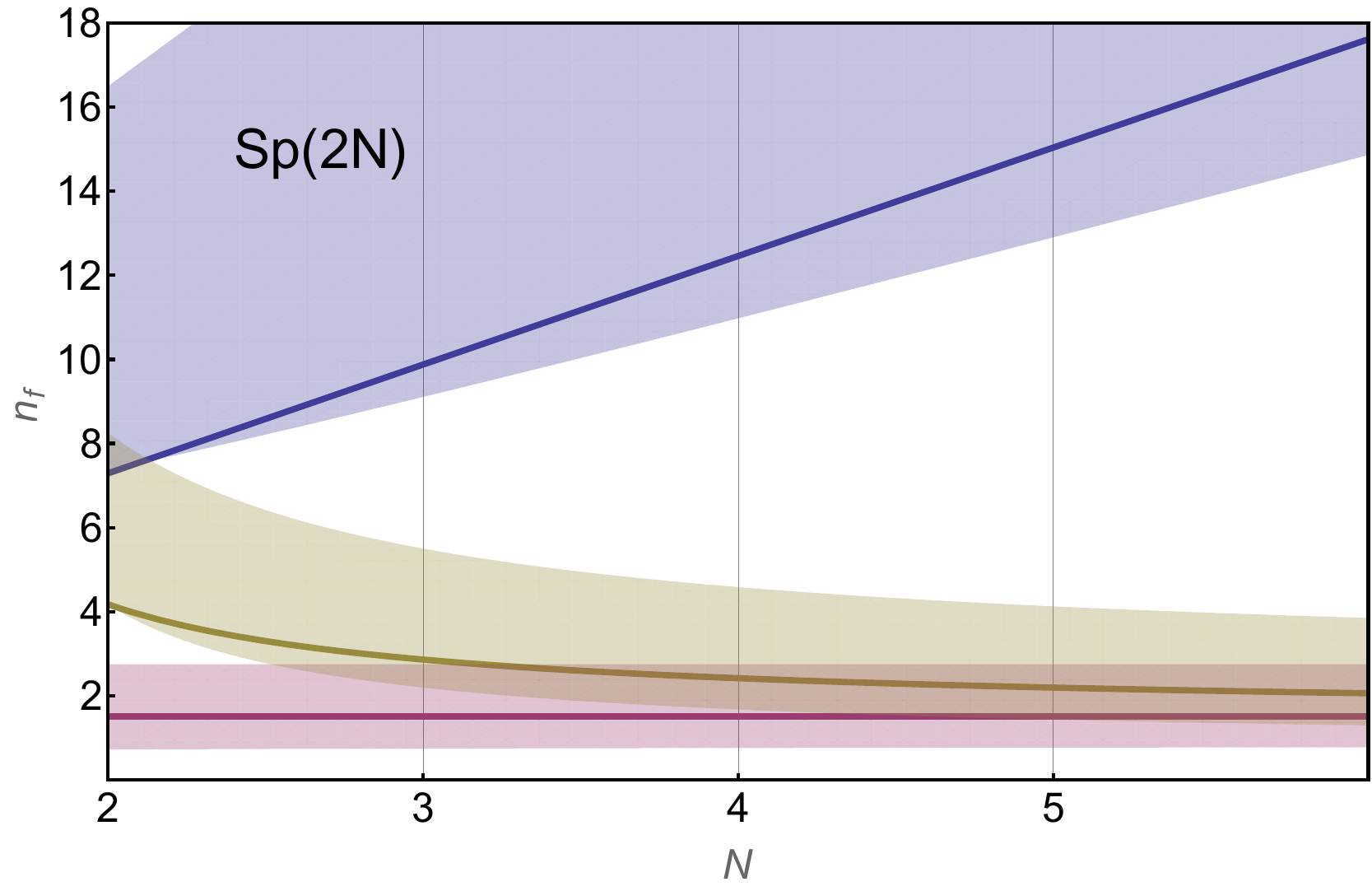}  
\caption{Conformal window for $SP(2N)$ groups for the fundamental representation (upper light-blue), two-index antisymmetric (next to the highest light-brown), two-index symmetric, i.e. the adjoint,  (bottom window in pink).}    
\label{SP-4loop-PD}
\end{center}
\end{figure*}

We summarize the results for the $SU(N)$ gauge groups in Fig.~\ref{4loop-PD} for the fundamental, two-index symmetric, two-index antisymmetric and adjoint representation. 
The conformal window at the four-loop level is considerably wider, for any representation, when compared with the Schwinger-Dyson results \cite{Sannino:2004qp,Dietrich:2006cm} or the one obtained using the critical number of flavors where the free energy changes sign, as suggested in \cite{Mojaza:2010cm}. 
For completeness we also show the conformal window for the orthogonal and symplectic gauge groups respectively in Fig.~\ref{SO-4loop-PD} and Fig.~\ref{SP-4loop-PD}. There is an universal trend towards the widening of the conformal regions with respect to earlier estimates using other nonperturbative methods. 

\subsection{All-orders beta function comparison}
We have recently demonstrated the existence of a scheme in which the all-orders beta function is \cite{Pica:2010mt}:
\begin{equation}
\frac{\beta(a)}{a} = - \frac{a}{3} \frac{11 C_A - 2 T_F\, n_f \, (2 +  \Delta_F \gamma)  }{1 - 2a \frac{17}{11}C_A} \ ,
\end{equation}
with 
\begin{equation}
\Delta_F = 1 + \frac{7}{11} \frac{C_A}{C_F} \ .
\label{eq:deltar}
\end{equation}
The {\it scheme independent} analytical expression of the anomalous dimension of the mass at the IR positive zero is\footnote{The same anomalous dimension was  guessed in \cite{Antipin:2009dz}.}:
 \begin{equation}
\gamma = \frac{11 C_A - 4 T_F n_f}{2 n_f T_F  \left( 1 + \frac{7}{11} \frac{C_A}{C_F} \right)} \ .
\label{eq:gammar}
 \end{equation}
 We  plot, for reference, in Figs.~\ref{4loop-PD}, \ref{SO-4loop-PD}, \ref{SP-4loop-PD} the lines corresponding to this anomalous dimension equal to unity. These are the solid thick curves  for the different representations. These lines could be viewed as the lower boundary of the conformal window if it is marked by the anomalous dimension to be unity. This value is also near to what seems to be consistent via gauge dualities \cite{Sannino:2009qc,Sannino:2009me} where the value of the anomalous dimension computed via \cite{Ryttov:2007cx} is corrected has shown in \cite{Pica:2010mt}. Important applications of duality to compute physical correlators such as the {\it conformal} S parameter \cite{Sannino:2010ca} has been proposed in \cite{Sannino:2010fh,DiChiara:2010xb}. 
 
\subsection{Four-loop anomalous dimensions}
\begin{figure}[t!]
\includegraphics[width=\columnwidth]{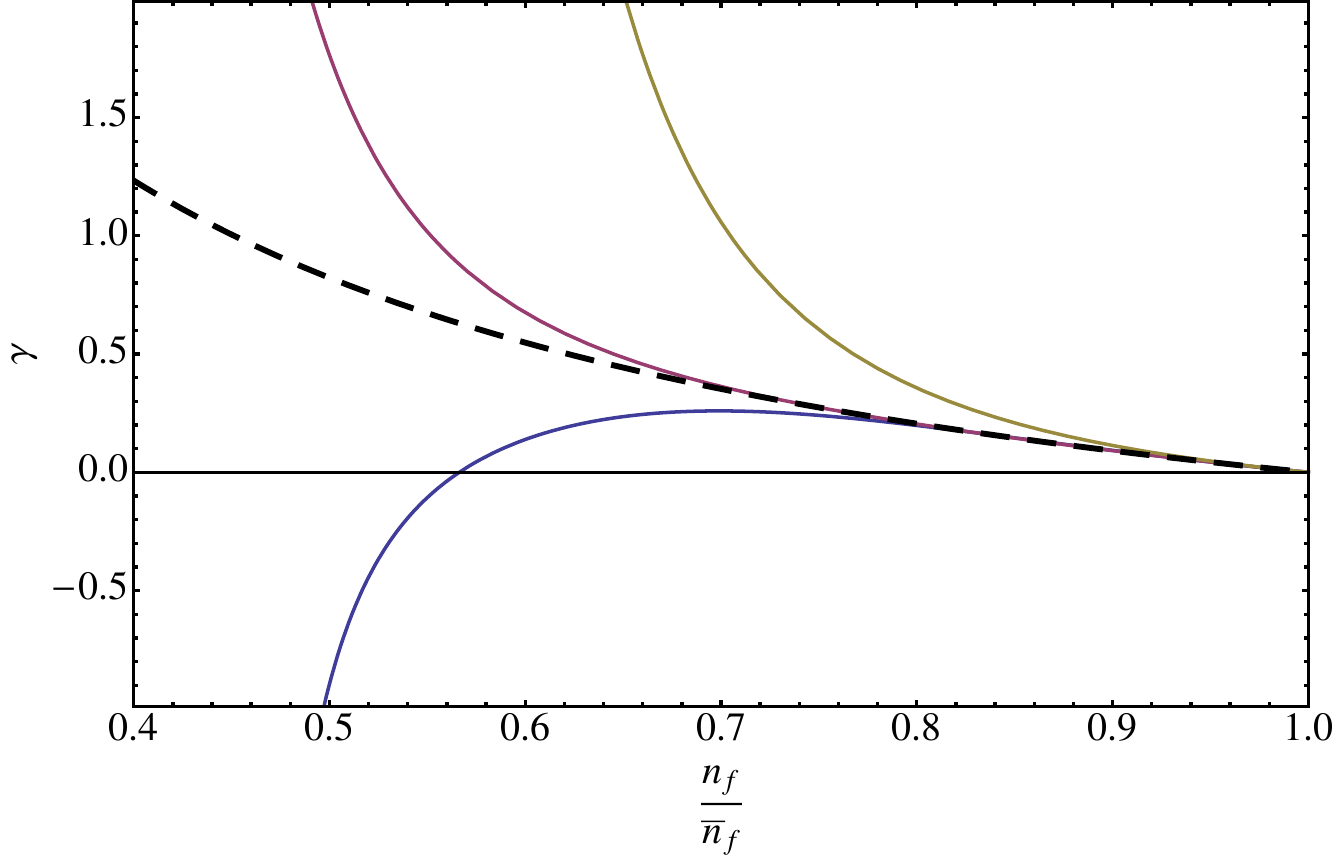}  
\caption{Anomalous dimension of the mass, at the infrared fixed point, for $SU(3)$ as function of the number of fundamental flavors at two loops (upper brown-curve), three-loops  (second curve from the top in magenta), all-orders (dashed-curve in black),  four-loops   (bottom curve in blue).}    
\label{gammafiga}
\end{figure}

We plot, for illustration, the anomalous dimension of the mass for the $SU(3)$ gauge theory, as function of the number of fundamental flavors, at the IR positive zero in Fig.~\ref{gammafiga}. The three solid lines correspond respectively, from top to bottom, to the two-, three- and four-loop results.
Perturbation theory is reliable only in a small range of flavors near $\overline{n}_f$. A similar behavior is observed for any other gauge group, matter representation and different number of colors. We note that the perturbative analysis of the anomalous dimension  appeared in \cite{Ryttov:2010iz}, while this paper was being completed.  There it has also been noted that the anomalous dimension,  to this order in perturbation theory, is smaller than for the three and two-loop case.  

Having at hand an all-order scheme-independent result, we compare it with the perturbative one.  The dashed line, in  Fig.~\ref{gammafiga}, is the all-order anomalous dimension  from \eqref{eq:gammar}. It is striking that the all-order result is much more well behaved than the four-loop predictions which, in this example, reach large and negative values long before loosing the IR positive zero.

\begin{figure}[ht]
\includegraphics[width=\columnwidth]{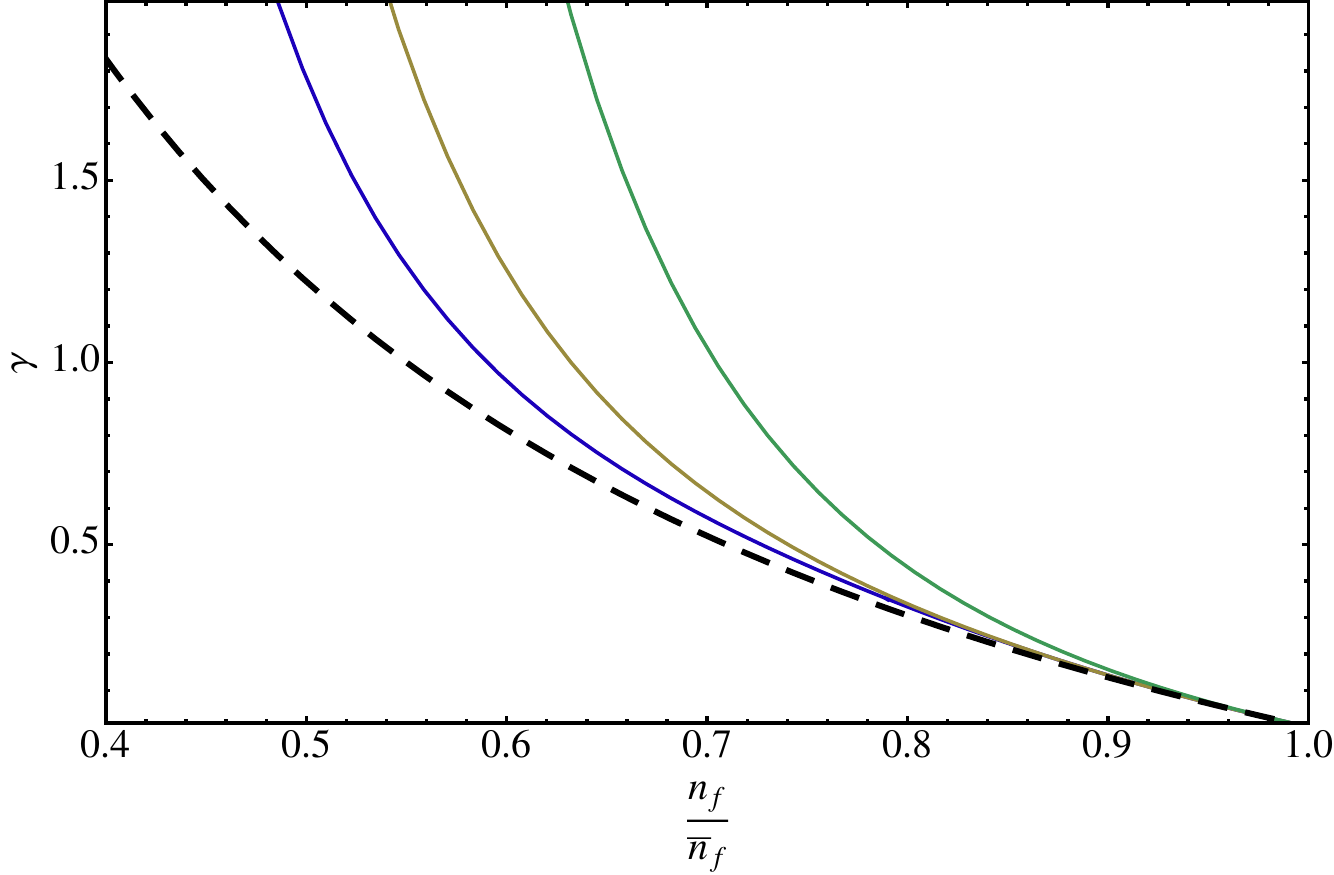}  
\caption{Anomalous dimension of the mass, at the infrared fixed point, for $SU(2)$ as function of the number of  adjoint Dirac flavors at two loops (up green-curve), three-loops  (second curve from the top),  four-loops (third curve in blue), all-orders (dashed curve in black). }    
\label{gammafig}
\end{figure}

Due to the phenomenological interest in models of minimal walking technicolor \cite{Sannino:2004qp,Hong:2004td,Dietrich:2005jn} we report the anomalous dimension at the fixed point also for the $SU(2)$ gauge theory with two-adjoint fermions in Fig.~\ref{gammafig}. 
These theories are being subject to intensive numerical investigations via lattice simulations \cite{Catterall:2007yx, DelDebbio:2008wb,Shamir:2008pb,Deuzeman:2008sc,DelDebbio:2008zf,Catterall:2008qk,DelDebbio:2008tv,DeGrand:2008kx,Hietanen:2008mr,Appelquist:2009ty,Hietanen:2009az,Deuzeman:2009mh,Hasenfratz:2009ea,DelDebbio:2009fd,Fodor:2009wk,Fodor:2009ar,DeGrand:2009hu,Catterall:2009sb,Bursa:2009we,Bilgici:2009kh,Kogut:2010cz,Hasenfratz:2010fi,DelDebbio:2010hu,DelDebbio:2010hx,Catterall:2010du}.

\section{Asymptotic safety  at large $\mathbf n_f$} 
\label{salvo}
To the four-loop order a positive UV zero appears for a sufficiently large number of flavors.  We have already observed that the value of the zero as function of number of flavors   decreases monotonically as $n_f^{-{2}/{3}}$ at four loops. In fact, it is possible to generalize this behavior to any finite order in perturbation theory. Consider the equation for the zeros of the beta function in which the leading powers in the number of flavors are made explicit: 
\beq
b_0 n_f + \sum_{k=1}^{\infty}b_k \, n_f ^k \,\alpha^k = 0 \ ,
\eeq
where  $b_0 = \beta_0/n_f$ and $b_k = \beta_k/n_f^k$.  We used the fact that first and second coefficient of the beta function are linear in the number of flavors and, in general, the successive coefficients have one extra power of $n_f$ \cite{Holdom:2010qs} and therefore the coefficients $b_k$ are finite at large number of flavors.

We define:  
\beq
x = n_f \alpha \ ,
\eeq
and the equation at any fixed perturbative order $P$ reads:  
\beq
b_0 n_f + \sum_{k=1}^{P}b_k \, x^k = 0 \ . 
\label{sommettina}
\eeq
{}At large $n_f$ the solution approaches: 
\beq
x = \left( - \frac{b_0 n_f}{b_P}\right)^{\frac{1}{P}}   \longrightarrow \alpha = \left( - \frac{b_0}{b_P}\right)^{\frac{1}{P}} {n_f}^{\frac{1-P}{P}} \ .
\eeq
There are $P$ complex solutions on a circle. A positive solution exists only if  $b_P$ is positive at large $n_f$. This is indeed the case, at the four-loop order, for any gauge theory  
showing that the UV positive zero vanishes as $n_f^{-2/3}$. 
If this UV zero persists to higher orders its location will change albeit will vanish faster as a function of $n_f$ when increasing $P$. 

It is, however, possible to sum exactly the perturbative infinite sum for the beta function, at large of number of flavors given that the leading coefficients are known. The result is: 
\beq
\frac{3}{4n_f T_F}\frac{\beta(a)}{a^2} = 1 + \frac{H(x)}{n_f} + {\cal{O}}\left(n_f^{-2}\right) \ . 
\eeq
The explicit form of $H(x)$ can be found  in \cite{Holdom:2010qs}. The important feature, here, is that $H(x)$  possess a negative singularity at $x = 3\pi /T_F$. This demonstrates that there always is a solution for the existence of a nontrivial UV fixed point at the leading order in $n_f$ for the following positive value of the coupling: 
\beq
\alpha_{\rm UV} = \frac{3\pi}{T_F n_f} \ . 
\eeq 
The function $H(x)$ has also other singularities which might signal the presence of new zeros which we will not consider here, but that would worth exploring.

Higher order terms in $n_f^{-1}$ can, in principle, modify the result if the singularity structure is such to remove or modify its location. 

A more complete discussion  of the singularity structure of the coefficients of the $n_f^{-1}$  expansion has appeared in \cite{Holdom:2010qs} also for QED.  It seems plausible that the smallest UV fixed point is an all-orders feature.  

\section{Conclusions} 
We presented the general features of the phase diagram for any gauge theory with  fermions transforming according to distinct representations of the underlying gauge group, at the four-loop order. The topology of the zeros of the perturbative beta function has been investigated. We discovered that only four distinct topologies are sufficient to classify the gauge dynamics of any theory.  

 At the IR stable fixed point, for positive values of the $\alpha$ coupling, we computed the anomalous dimensions. We have also shown that these are well described by the all-orders beta function for any theory. 

Finally, by investigating the large $n_f$ limit we argued the possible existence, to all orders, of a nontrivial UV fixed point for any non-Abelian gauge theory at large number of flavors.

\acknowledgments
We thank the CERN Theory Institute for its kind hospitality during the meeting {\it Future directions in lattice gauge theory - LGT10}, 25 - July to 6 - August  2010 where this work started with the discovery that the Ryttov-Sannino beta function had to be corrected \cite{Pica:2010mt} leading to smaller anomalous dimensions. We discussed the findings with several scientists since.
Part of the results presented in this manuscript were used in \cite{Mojaza:2010cm} to compute the free-energy to the order $g^6 \ln g$.  

While this paper was being finalized the related paper \cite{Ryttov:2010iz} by Ryttov and Shrock appeared which partially overlaps with the present work.

Finally, we are happy to thank Oleg Antipin,  Chris Kouvaris, Matin Mojaza, Marco Nardecchia, Thomas Ryttov and Ulrik I. S\o ndergaard for useful discussions.

{\it Note added: In the present version we correct the group theory factors  $d_F d_A$ and $d_F d_F$ of Table II for the SO and Sp groups that enter in the related four loop beta functions. Consequently we updated figures 3 and 4, as well as Table I. None of our main points are affected. We stress that we had already employed the correct group factors in a follow up paper \cite{Hietanen:2012sz} where we reported the adjourned four-loop conformal window for the SO group. The latter now matches the one presented here. We thank T.~Ryttov for reminding us to update the four loops invariants of the present work.}

\appendix
\section{Group factors and  perturbative coefficients}
The four-loop beta function coefficients are \cite{vanRitbergen:1997va}:
\label{beta-a}
\renewcommand{\arraystretch}{1.3}
\begin{widetext}
\begin{eqnarray} 
\label{eq:beta3}
\beta_{0} & = &  \frac{11}{3} C_A - \frac{4}{3} T_F n_f  \ ,
  \\
 \beta_{1} & = &
 \frac{34}{3}C_A^2 - 4 C_F T_F n_f -\frac{20}{3} C_A T_F n_f  \ ,\nonumber \\
 \beta_{2} & = &  \frac{2857}{54} C_A^3
 +2 C_F^2 T_F n_f - \frac{205}{9} C_F C_A T_F n_f  - \frac{1415}{27} C_A^2 T_F n_f
 + \frac{44}{9} C_F T_F^2 n_f^2
  + \frac{158}{27} C_A T_F^2 n_f^2  \ ,
   \nonumber \\
\beta_{3} & = &
    C_A^4 \left( \frac{150653}{486} - \frac{44}{9} \zeta_3 \right)   
    +  C_A^3 T_F n_f  
      \left(  - \frac{39143}{81} + \frac{136}{3} \zeta_3 \right)
    + C_A^2 C_F T_F n_f 
\left( \frac{7073}{243} - \frac{656}{9} \zeta_3 \right)
   +     C_A C_F^2 T_F n_f 
      \left(  - \frac{4204}{27} + \frac{352}{9} \zeta_3 \right)
\nonumber \\ & &
   + 46 C_F^3 T_F n_f 
   +  C_A^2 T_F^2 n_f^2 
      \left( \frac{7930}{81} + \frac{224}{9} \zeta_3 \right)
    +  C_F^2 T_F^2 n_f^2 
      \left( \frac{1352}{27} - \frac{704}{9} \zeta_3 \right)
    +  C_A C_F T_F^2 n_f^2 
      \left( \frac{17152}{243} + \frac{448}{9} \zeta_3 \right)
  + \frac{424}{243} C_A T_F^3 n_f^3 \nonumber\\ & &
   + \frac{1232}{243} C_F T_F^3 n_f^3  
       +  \frac{d_A^{a b c d}d_A^{a b c d}}{N_A }  
             \left(  - \frac{80}{9} + \frac{704}{3} \zeta_3 \right)
       + n_f \frac{d_F^{a b c d}d_A^{a b c d}}{N_A }  
            \left(   \frac{512}{9} - \frac{1664}{3} \zeta_3 \right)
       + n_f^2 \frac{d_F^{a b c d}d_F^{a b c d}}{N_A }   
            \left(  - \frac{704}{9} + \frac{512}{3} \zeta_3 \right)  \ .
 \label{mainbeta} \nonumber 
\end{eqnarray} 
\end{widetext}
Here $\zeta_x$ is the Riemann zeta-function evaluated at $x$, $T_F^a$ with $a=1,\ldots, N_F$ are the generators for a generic representation $F$ with dimension $N_F$.  The generators are normalized via tr$(T^a_F T^b_F) = T_F \delta^{a b}$ and the quadratic Casimirs are  $[T^a_F T^a_F]_{ij} = C_F \delta_{ij}$. The subscript $A$ refers to the adjoint representation in the formulae in the text. Here the number of fermions is indicated by $n_f$.

The symbols $d_F^{abcd}$ are the forth-order 
group invariants expressed in terms of contractions between the following fully 
symmetrical tensors:
\begin{eqnarray}
 d_F^{a b c d} & = & \frac{1}{6 } {\rm Tr }  \left[
   T^a_F T^b_F T^c_F T^d_F
 + T^a_F T^b_F T^d_F T^c_F 
 + T^a_F T^c_F T^b_F T^d_F  \right. \nonumber \\
 & & \left. \hspace{4mm}
 + T^a_F T^c_F T^d_F T^b_F 
 + T^a_F T^d_F T^b_F T^c_F 
 + T^a_F T^d_F T^c_F T^b_F  
  \right] \hspace{1mm}
\end{eqnarray}
For readear's convenience we provide in table \ref{coefficienti} the relevant group factors. 

The coefficients of the anomalous dimension to four-loops are \cite{Vermaseren:1997fq}:
 \renewcommand{\arraystretch}{1.3}
\begin{widetext}
\begin{eqnarray} 
 \label{maingamma}
\gamma_{0} & = &
 3 C_F
 \\
 \gamma_{1} & = &
 \frac{3}{2}C_F^2+\frac{97}{6} C_F C_A -\frac{10}{3} C_F T_F n_f 
  \nonumber \\
 \gamma_{2} & = &
  \frac{129}{2} C_F^3 - \frac{129}{4}C_F^2 C_A
 + \frac{11413}{108}C_F C_A^2 
 +C_F^2 T_F n_f (-46+48\zeta_3)
+C_F C_A T_F n_f \left( -\frac{556}{27}-48\zeta_3 \right)  
- \frac{140}{27} C_F T_F^2 n_f^2 
   \nonumber \\
\gamma_{3} & = & 
  C_F^4 \left(-\frac{1261}{8} - 336\zeta_3 \right)
   + C_F^3 C_A \left( \frac{ 15349}{12} + 316 \zeta_3 \right)
   + C_F^2 C_A^2 \left(-\frac{ 34045}{36} - 152 \zeta_3 + 440\zeta_5 \right)
   + C_F C_A^3 \left( \frac{70055}{72} + \frac{1418}{9} \zeta_3
                      - 440 \zeta_5 \right)
 \nonumber \\ & &
  + C_F^3 T_F n_f \left( -\frac{280}{3} + 552 \zeta_3 - 480 \zeta_5 \right)
  + C_F^2 C_A T_F n_f \left(- \frac{8819}{27} + 368 \zeta_3 
                            - 264 \zeta_4 + 80 \zeta_5 \right)
 \nonumber \\ & &
  + C_F C_A^2 T_F n_f \left(- \frac{65459}{162} 
                  - \frac{2684}{3} \zeta_3 + 264 \zeta_4
                    + 400 \zeta_5 \right)
  + C_F^2 T_F^2 n_f^2 \left( \frac{304}{27} - 160 \zeta_3 
                            + 96 \zeta_4 \right)
 \nonumber \\ & &
  + C_F C_A T_F^2 n_f^2 \left( \frac{1342}{81} 
                             + 160 \zeta_3 - 96 \zeta_4 \right) 
  + C_F T_F^3 n_f^3 \left(- \frac{664}{81} + \frac{128}{9} \zeta_3 \right)
  + \frac{d_F^{a b c d}d_A^{a b c d}}{N_F}   
            \left(- 32 + 240 \zeta_3  \right)
  +  n_f \frac{d_F^{a b c d}d_F^{a b c d}}{N_F}   
            \left( 64 - 480 \zeta_3  \right) 
 \nonumber  
 \end{eqnarray}
\end{widetext}

The results of  \eqref{eq:beta3}and \eqref{maingamma}
ares valid for an arbitrary semi-simple compact Lie group. 
The result for QED (i.e. the group U(1))
is included in Eq. (\ref{maingamma}) by substituting
$C_A = 0$, $d_A^{a b c d} = 0$, $C_F = 1$, $T_F = 1$, $(d_F^{a b c d})^2=1$,
$N_F = 1$.

These coefficients were obtained in an arbitrary
covariant gauge for the gluon field and are gauge independent.
 
 \begin{table*}
 \begin{equation*}
\begin{array}{c|c|c|c|c|c}
Rep. &N_F & T_F  &C_F & d_F^{abcd} d_A^{abcd}/N_F & d_F^{abcd} d_F^{abcd}/N_F \\\hline \hline
\multicolumn{6}{c}{SU(N)}\\\hline \hline
 FUND&N & \frac{1}{2} & \frac{N^2-1}{2 N} & \frac{1}{48} (N-1) (N+1) \left(N^2+6\right) & \frac{(N-1) (N+1) \left(N^4-6
   N^2+18\right)}{96 N^3} \\
 ADJ& N^2-1 & N & N & \frac{1}{24} N^2 \left(N^2+36\right) & \frac{1}{24} N^2 \left(N^2+36\right) \\
 2-SYM&\frac{1}{2} N (N+1) & \frac{N+2}{2} & N-\frac{2}{N}+1 & \frac{1}{24} (N-1) (N+2) \left(N^2+6 N+24\right) & \frac{(N-1) (N+2)
   \left(N^5+14 N^4+72 N^3-48 N^2-288 N+576\right)}{48 N^3} \\
 2-ASY&\frac{1}{2} (N-1) N & \frac{N-2}{2} & N-\frac{2}{N}-1 & \frac{1}{24} (N-2) (N+1) \left(N^2-6 N+24\right) & \frac{(N-2) (N+1)
   \left(N^5-14 N^4+72 N^3+48 N^2-288 N-576\right)}{48 N^3}\\ \hline\hline
\multicolumn{6}{c}{SO(N)}\\\hline \hline
FUND & N & 1 & \frac{N-1}{2} & \frac{1}{48} (N-2) (N-1) \left(N^2-7 N+22\right) & \frac{1}{48} (N-1) \left(N^2-N+4\right) \\
ADJ & \frac{1}{2} (N-1) N & N-2 & N-2 & \frac{1}{24} (N-2) \left(N^3-15 N^2+138 N-296\right) & \frac{1}{24} (N-2) \left(N^3-15 N^2+138 N-296\right) \\
2-SYM & \frac{1}{2} \left(N^2+N-2\right) & N+2 & N & \frac{1}{24} (N-2) N \left(N^2-N+28\right) & \frac{1}{24} N \left(N^3+13 N^2+110 N+104\right) \\ \hline\hline
\multicolumn{6}{c}{SP(2N)}\\\hline \hline
FUND & 2 N & \frac{1}{2} & \frac{1}{4} (2 N+1) & \frac{1}{192} (N+1) (2 N+1) \left(2 N^2+7 N+11\right) &
\frac{1}{384} (2 N+1) \left(2 N^2+N+2\right) \\
ADJ & N (2 N+1) & N+1 & N+1 & \frac{1}{48} (N+1) \left(2 N^3+15 N^2+69 N+74\right) & \frac{1}{48} (N+1)
\left(2 N^3+15 N^2+69 N+74\right) \\
2-ASY & N (2 N-1)-1 & N-1 & N & \frac{1}{48} N (N+1) \left(2 N^2+N+14\right) & \frac{1}{48} N \left(2
N^3-13 N^2+55 N-26\right) \\ \hline\hline
 \end{array}
  \end{equation*}
\caption{Relevant group factors for $SU(N)$, $SO(N)$ and $SP(2N)$.  \label{coefficienti}}
\end{table*}


\end{document}